\documentclass[preprint,showpacs,preprintnumbers,amsmath,amssymb]{revtex4}

\usepackage{epsfig}
\usepackage{graphics}
\usepackage{graphicx}
\usepackage{dcolumn}
\usepackage{bm}

\begin{document}

\preprint{APS/123-QED}

\title{Experimental realization of
strange nonchaotic attractors in a quasiperiodically forced electronic
circuit}

\author{K.~Thamilmaran$^1$}%
\author{D.~V.~Senthilkumar$^1$}
\author{A.~Venkatesan$^2$}%
\author{M.~Lakshmanan$^1$}%
 \email{lakshman@cnld.bdu.ac.in}
\affiliation{%
$^1$Centre for Nonlinear Dynamics,Department of Physics,
Bharathidasan University, Tiruchirapalli - 620 024, India\\
}%
 \affiliation{%
 $^2$Department of Physics,
 Nehru Memorial College, Puthanampatti - P.O,
 Tiruchirappalli - 621 007, India \\
 }%

\date{\today}

\begin{abstract}  
We have identified the three prominent routes, namely Heagy-Hammel,
fractalization and intermittency routes, and their mechanisms for the birth of
strange nonchaotic attractors (SNAs) in a quasiperiodically forced electronic
system constructed using a negative conductance series LCR circuit with a diode
both numerically  and experimentally.  The birth of SNAs by these three routes
is verified from both experimental and their corresponding numerical data by maximal Lyapunov
exponents, and their variance, Poincar\'e maps, Fourier amplitude spectrum,
spectral distribution function  and finite-time Lyapunov exponents. Although 
these three routes have been identified numerically  in different dynamical
systems, the experimental observation of all these mechanisms is reported for the 
first time to our knowledge and that too in a single second order electronic
circuit. 
\end{abstract}

\pacs{05.45.+b}
\maketitle

\section{\label{sec:level1}Introduction}

Strange nonchaotic attractors are regarded as structures in  between regularity
and chaos. They are geometrically strange as evidenced by their fractal nature,
which is common to all chaotic systems. However, they are nonchaotic in a
dynamical sense because they do not  show sensitivity with respect to changes in
initial conditions (as evidenced by negative Lyapunov exponents), just like,
regular systems. Following the initial study of Grebogi et al.
~\cite{cgeo1984}, several theoretical as well as experimental studies have been
made pertaining  to the existence and characterization of SNAs in different
quasiperiodically  driven nonlinear dynamical systems. In particular the SNAs have
been reported to arise in many physically relevant situations such as
the quasiperiodically forced pendulum~\cite{fjreo1987}, the quantum particles in
quasiperiodic potentials~\cite{abeo1985}, biological
oscillators~\cite{mdcg1989}, the quasiperiodically driven Duffing-type
oscillators~\cite{jfhwld1991,tyycl1996,tkjw1993,avml2000}, velocity dependent
oscillators~\cite{avml1997}, electronic
circuits~\cite{tykb1997,zlzz1996,avkm1999} and in certain
maps~\cite{apvm1997,aspuf1995,aspuf1994,vsatkv1996,tnkk1996,avml2001,brheo2001,
sykwl2003,wlsyk2004,jfhsmh1994}.    Also, these exotic attractors were
confirmed by an experiment consisting of a quasiperiodically forced, buckled,
magneto-elastic ribbon~\cite{wldmls1990}, in analog simulations of a
multistable potential~\cite{tzfm1992}, and in a neon glow discharge
experiment~\cite{wxdhd1997}. The SNAs are also related to the Anderson localization
in the Schr$\ddot {\text o}$dinger equation with a quasiperiodic
potential~\cite{jakis1997,aprr1999} and they may have a practical application
in secure communication~\cite{csztlc1997,rr1997,rcamg2002}.

The existence of SNAs in the above physically relevant systems has naturally 
motivated further intense investigations on their nature and occurrence. A 
question  of intense further  interest is the way in which they arise and 
ultimately become chaotic.   In this context, several routes have been
identified  in recent times and for a few of them  typical mechanisms have been
found for the creation of SNAs. The major routes by which the SNAs appear may
be broadly classified as follows: torus doubling route to chaos via
SNAs~\cite{jfhsmh1994},  gradual fractalization of torus~\cite{tnkk1996},  the
appearance of SNAs via blowout bifurcation~\cite{tyycl1996}, the occurrence of
SNAs  through intermittent
phenomenon~\cite{avkm1999,apvm1997,brheo2001,sykwl2003,wlsyk2004,sykwl2004},
the formation of SNAs via homoclinic collision~\cite{aprr1999}, remerging of
torus doubling bifurcations and the birth of SNAs~\cite{avml1997}, the
existence  of SNAs in the transition from two-frequency to three-frequency
quasiperiodicity~\cite{tkjw1993}, the transition from three-frequency
quasiperiodicity to chaos via a SNA~\cite{mdcg1989} and the transition to chaos
via strange nonchaotic trajectories on the torus~\cite{tkloc1997}. Different
mechanisms have been identified for some of the above routes, which are summarised
in Table I. 
\begin{table}
\caption{Routes and mechanisms of the onset of various SNAs}
\small
\begin{tabular}{|p{7cm}|p{7cm}|}
\hline
{\bf \emph \;\;\;\;\;\;\;\;\;\;\;\;\;\;Type of route} & {\bf \emph \;\;\;\;\;\;\;\;\;\;\;\;\;\;\;\;Mechanism}  \\
\hline
Heagy-Hammel \cite{jfhsmh1994}  & Collision of period-doubled torus with its unstable
parent\\
\hline
Gradual Fractilization \cite{tnkk1996} & Increased wrinkling of torus without any interaction
with nearby periodic orbits\\
\hline
On-off intermittency \cite{tyycl1996} &Loss of transverse stability of torus\\
\hline
Type-I intermittency \cite{apvm1997} & Due to saddle-node bifurcation, a torus
is replaced by SNA\\
\hline
Type-III intermittency \cite{avkm1999}  & Subharmonic instability\\
\hline
Homoclinic collision \cite{aprr1999} & Homoclinic collisions of the
quasiperiodic orbits\\
\hline
\end{tabular}
\end{table} 

Among these various routes/mechanisms for the birth of SNAs, the Heagy-Hammel,
the gradual fractalization and the intermittency routes/mechanisms to SNAs are quite
general and very robust to observe in a number of quasiperiodically forced
nonlinear dynamical systems.   So far, these dynamical transitions are
identified only through numerical analysis in different dynamical systems,
prominent among being discrete and continuous flow systems. Eventhough there
exist various experimental realizations of SNAs in physical systems
\cite{wldmls1990,tzfm1992,wxdhd1997}, the genesis of SNAs through different
routes/mechanisms have not yet been reported experimentally to the best of our
knowledge, except for type-III intermittent route by  two of the present
authors and K. Murali~\cite{avkm1999}.  In view of this fact, in the present work, we
consider a simple nonlinear electronic circuit system, a second-order
dissipative  nonautonomous negative conductance series LCR circuit, and
investigate the dynamics of the circuit under quasiperiodic forcing. We have
identified that the circuit exhibits the three familiar dynamical transitions,
namely  Heagy-Hammel, fractalization and intermittency transitions involving
SNAs.  Further, the dynamical transitions are characterized from both
experimental and their corresponding numerical data by   the maximal Lyapunov
exponents, and their variance, Poincar\'e maps, Fourier amplitude spectrum,
spectral distribution function  and finite-time Lyapunov exponents. We believe
that this is the first experimental demonstration of the existence of all the
three prominent routes/mechanisms to SNA and that too in a simple single
electronic circuit to the best of our knowledge.

The paper is organized as follows.  In Sec. II, we present a brief introduction
of the experimental realization of the  quasiperiodically forced negative
conductance series LCR circuit with diode.   In Sec. III, we describe the phase
diagram for the circuit where the  regions corresponding to the different
dynamical behaviors are delineated as a function of parameters based on numerical 
analysis. Section IV is
devoted to the computer simulation studies and  experimental confirmation of
the creation of strange nonchaotic attractors via Heagy-Hammel route while in
Sec. V the creation of SNAs through gradual fractalization
is studied both numerically and experimentally.  In Sec. VI, the type-I intermittent
route to SNA is shown to exist both numerically and experimentally. Finally,
in Sec. VII, we summarize our results.

\section{Circuit realization}
We consider here the simple second-order nonlinear dissipative nonautonomous
negative conductance series LCR circuit with a single voltage generator
introduced by us very recently
~\cite{ktdvs2005}  and shown in Fig.~1(a).
\begin{figure}
\centering
\includegraphics[width=0.75\columnwidth]{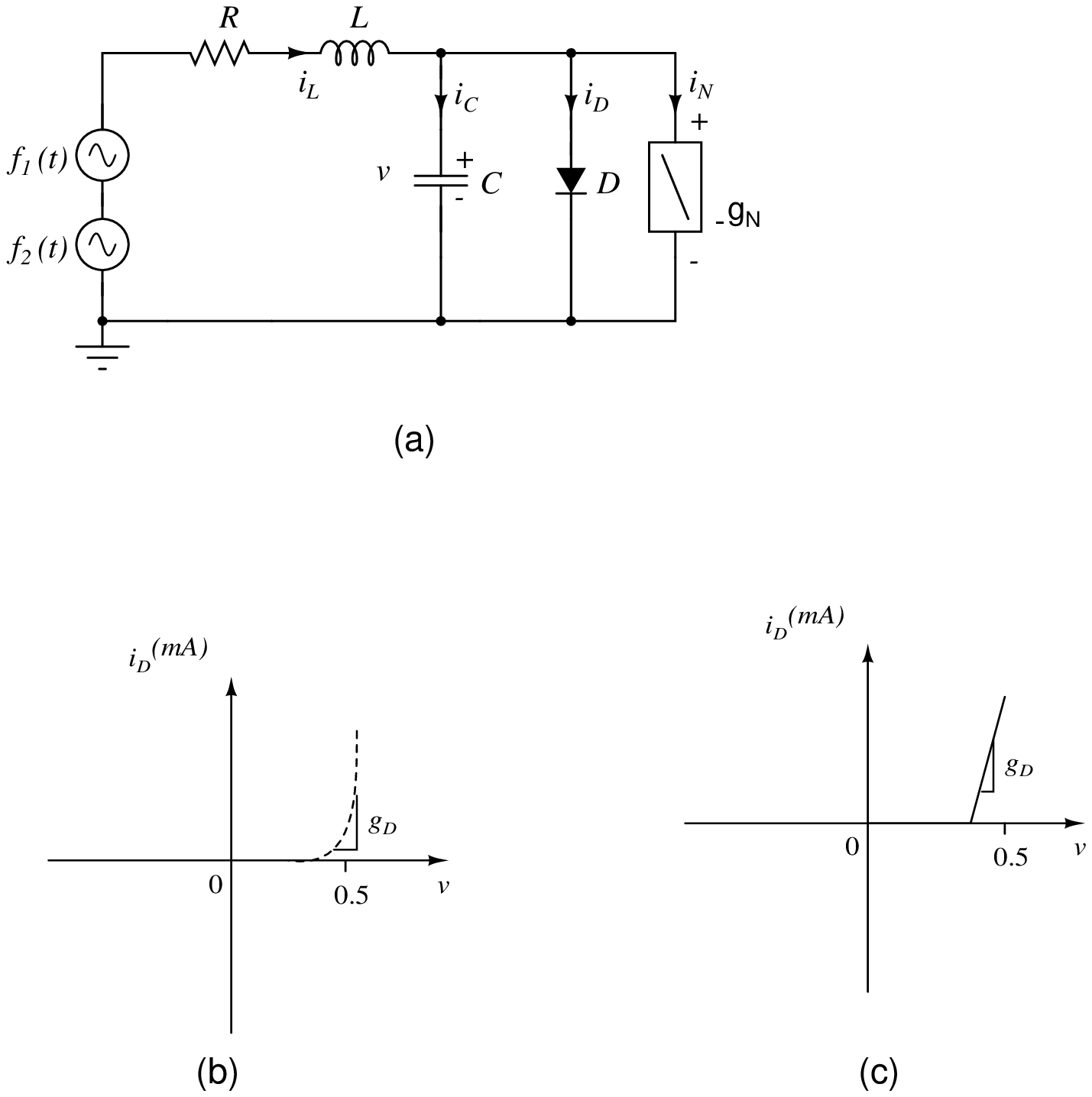}
\caption{\label{circuit}(a) Circuit realization of a simple nonautonomous
circuit.  Here, $D$ is the {\it{p-n junction}} diode, and $g_N$ is negative
conductance. The parameter values of the other elements are fixed as
$L=50.0~mH$, $C=10.32~nF$.  The external emf $f_1(t)=E_{f1} \sin \omega_{f1}t$ 
and $f_2(t)=E_{f2} \sin \omega_{f2}t$ are the function generators
$(HP~33120A)$.  The values of $\omega_{f1}$ and $\omega_{f2}$ are chosen as
$5982.0~Hz$ and $13533.0~Hz$ respectively. The forcing amplitude $E_{f2}$ is
fixed as 0.15 $V$.  The other forcing amplitude $E_{f1}$ and the resistance
$R$ are taken as control parameters  which are being varied in our analysis,
(b) $i-v$ characteristics of the {\it{p-n junction}} diode and (c) two segment piecewise-linear
function.}
\end{figure}
The circuit consists of a series LCR network, forced by two sinusoidal
voltage generators $f_1(t)$ and $f_2(t)$ (HP 33120A series).  Two extra
components, namely a {\it{p-n junction}} diode (D) and a linear negative
conductor $g_N$,
are connected in parallel to the forced series LCR circuit.  The negative
conductor used here is a standard op-amp based negative impedance
converter (NIC).  The diode operates as a nonlinear conductance, limiting the
amplitude of the oscillator.  In the Fig.~1(a), $v,i_L$ and $i_D$ denote the
voltage across the capacitor $C$, the current through the inductor $L$ and the
current through the diode $D$, respectively.  The actual $v-i$ characteristic of the diode
(given by Fig.~1(b)) is approximated  by the usual two segment piecewise-linear
function (see Fig.~1(c)) which facilitates numerical analysis considerably.  
The state equations governing the presently proposed
circuit given in Fig. 1  are a set of two first-order nonautonomous
differential equations:
\begin{subequations} 
\begin{eqnarray} 
C\frac{dv}{dt} &=&{i_L- i_D +g_{N}v}, \\
L\frac{di_L}{dt} &=&{-Ri_L-v + E_{f1} \sin( \omega_{f1} t ) + E_{f2} \sin( \omega_{f2} t)}.
\end{eqnarray} 
Here, 
\begin{eqnarray}
i_D(v)=
\left\{
\begin{array}{cc}
g_D(v-V),& v \ge V, \\
0,& v < V, \\
\end{array} \right.
\end{eqnarray}
\end{subequations} 
where $g_D$ is the slope of the characteristic curve of the diode, $E_{f1}$ and
$E_{f2}$ are the amplitudes and $\omega_{f1}$ and $\omega_{f2}$ are the angular
frequencies of the forcing functions $f_1(t)=E_{f1} \sin \omega_{f1}t$ and 
$f_2(t)=E_{f2} \sin\omega_{f2}t$, respectively.  In the absence of $E_{f2}$,
the circuit (Fig. 1(a)) has been shown to exhibit chaos and also strong chaos
not only through the familiar period-doubling route but also via torus
breakdown followed by period-doubling bifurcations~\cite{ktdvs2005}. In order
to construct the actual experimental circuit, the numerical simulation is used
to determine the correct parametric values for observing strange nonchaotic
attractor. The values of diode conductance $g_D$, negative conductance $g_N$ and 
break voltage $V$ are fixed as 1313~$\mu S$, $-$0.45~$m S$, and 0.5~$V$ respectively.
After some trial and error, we chose the actual experimental values of 
the inductance, $L$,
capacitance, $C$ and external frequencies $\omega_{f1}$ and $\omega_{f2}$ 
as 50~$mH$, 10.32~$nF$, 5892~$Hz$ and 13533~$Hz$.    

In order to study the dynamics of the circuit in detail, Eq. (1) can
be converted into a convenient normalized form for numerical analysis by using
the the following rescaled variables and parameters $\tau=t/{\sqrt{LC}}$,
~$x=v/V$, ~$y=(i_L/V) ( \sqrt(LC)$, ~$E_1=E_{f1}/V$, $E_2=E_{f2}/V$,
~$\omega_1=\omega_{f1} \sqrt(LC)$, ~$\omega_2=\omega_{f2} \sqrt(LC)$,
~$a=R\sqrt(C/L)$, ~$b=g_N\sqrt(L/C)$, and ~$c=g_D\sqrt(L/C)$.\\ The normalized
evolution equation so obtained is
\begin{subequations}
\begin{eqnarray}
\dot{x} & = & y + f(x),  \nonumber \\ 
\dot{y} & = & -x -ay + E_1 \sin( \theta)+E_2 \sin( \phi), \nonumber\\
\dot{\theta} & = & \omega_1, \nonumber\\
\dot{\phi} & = & \omega_2, 
\end{eqnarray}
where
\begin{eqnarray}
f(x)=
\left\{
\begin{array}{cc}
(b-c)x+c,& x \ge 1, \\
bx,& x < 1. \\
\end{array} \right.
\end{eqnarray}
\end{subequations}
Here dot stands for the differentiation with respect to $\tau$.

The dynamics of Eq. (2) now depends on the parameters $a$, $b$, $c$,
$\omega_1$, $\omega_2$, $E_1$ and $E_2$. The rescaled parameters correspond to the
values $b=0.99051$, $c=2.89$, $\omega_1=0.133841$, $\omega_2=0.307411$ and
$E_2=0.3$. The amplitude of external quasiperiodic forcing $E_1$ and  the value
of  $a$ (or equivalently $E_{f1}$ and $R$ in Eq.~(1)) are taken as control parameters which
are being varied in our numerical  (and experimental) studies.

\section{Two parameter Phase diagram}
To be concrete, we first consider the dynamics of the system (2) and
numerically integrate it. Using various characteristic quantities such as
Lyapunov exponents, power spectral measures and distribution of finite-time
Lyapunov exponents, we distinguish periodic, quasiperiodic, strange nonchaotic
and chaotic attractors.   In particular, the Poincar\'e surface of section plot
in the $(\phi-x)$ plane with $ \phi$ modulo $2 \pi$  can clearly indicate
whether an attractor possesses a geometrically smooth or complicated
structure.  However, the estimation of the Lyapunov exponents for this
attractor(that is positive or negative value including zero) as well as its
variance  will identify whether it is a chaotic or nonchaotic one.  In addition
to the fact that the Lyapunov exponents are negative for SNAs, the variance - the
fluctuations in the measured value of the Lyapunov exponents on SNAs - is also
found to be large.  Finer distinction among SNAs formed via different
mechanisms can be  made by analyzing the nature of the variation of Lyapunov
exponents and its  variance near the transition values of the control
parameters.  Then we experimentally confirm the results for circuit Eq.~(1)
geometrically by observing the phase trajectory and the power spectrum.  For our
experimental study of the circuit given in Fig. 1, a two dimensional projection
of the attractor is obtained  by measuring the voltage $v$ across the capacitor
$C$ and the current $i_L$ through the inductor $L$  and connected to the $X$
and $Y$ channels of an oscilloscope. The phase trajectory obtained in the
experiment is compared  with the numerical trajectory. Then, a live picture of
the corresponding power spectrum (obtained from a digital storage oscilloscope
- HP 54600 series) of the projected attractor has also been used to distinguish
the different attractors.   In particular, to quantify the changes in the power
spectrum obtained  by numerically and experimentally, we compute the so-called
spectral distribution function $N(\sigma)$,which is defined to be the number of peaks
in the Fourier amplitude spectrum larger than some value, say $\sigma$. Scaling
relations have been identified in the form  $N(\sigma)$=$\log_{10}(1/ \sigma)$ for the
case of two-frequency quasiperiodic attractors and   $N(\sigma)=$~
$\sigma^{-\beta}$, $1<\beta<2$,  for the strange nonchaotic  attractors.
 
Further to identify the different attractors in the two-parameter plane the
dynamical transitions are traced out by two scanning procedures, both
numerically and experimentally: (1) varying $E_1$(or $E_{f1}$) at fixed $a$
($= R\sqrt(C/L)$, and (2) varying $a$ (or $= R$) at fixed $E_1$(or
$E_{f1}/V$) in a 1000 X 1000 grid.  The resulting phase diagram in the
$(a-E_1)$ parameters space in the region $a\in$(0.9, 0.98) and  $E_1\in$(0.34, 0.7)
is shown in Fig. 2 which has also been verified in the corresponding
$(R-E_{f1})$ experimental parameter space. The various features indicated in
the phase diagram are summarized and the main interesting features of the 
dynamical transitions are elucidated in the following.
\begin{figure} 
\centering
\includegraphics[width=0.9\columnwidth]{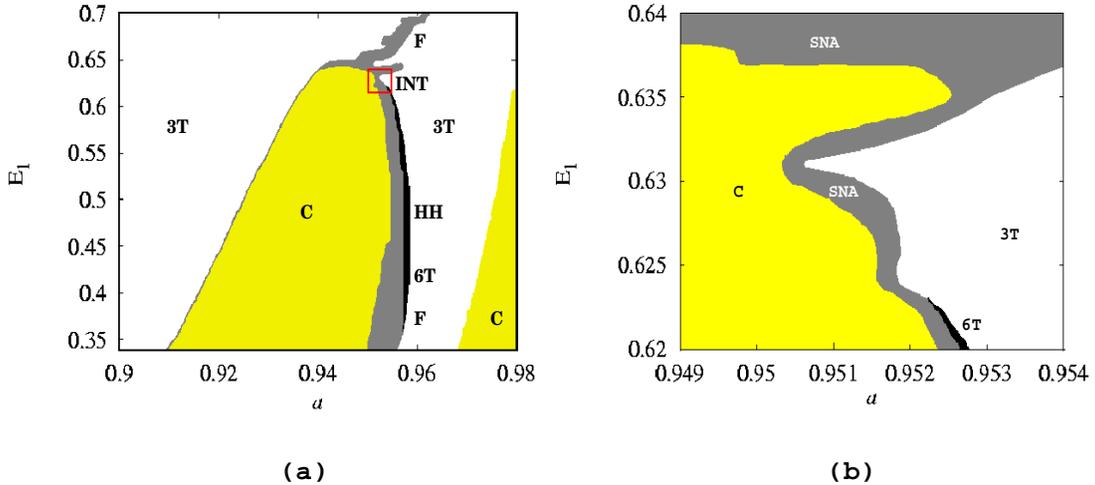} 
\caption{\label{2phase}(Color online)(a) Phase diagram in the $(a-E_1)$ plane for the circuit given in Fig. 1, represented by Eq. (2) and obtained from numerical data.
3T and 6T correspond to torus of period-3 and period-6 attractors, respectively. 
F, HH and INT denote the formation of SNAs through gradual fractalization, Heagy-Hammel and intermittency 
routes, respectively. C corresponds to chaotic attractor. (b) An enlarged version of the
intermittent region indicated in (a).}   
\end{figure} 

Transitions from the right to left lower down in the ($a,E_1$) space, through
fractalization  of the period-3 (3T) quasiperiodic attractors to SNA
and then to chaos, occur for 0.953 $< a <$ 0.955 and 0.35 $<E_1<$ 0.37. It is
denoted as {\bf F} in Fig. 2.  

Moving from right to left in the middle region, one finds a torus doubling bifurcation 
from a period-3 torus (3T) to  a period-6 (6T) quasiperiodic attractor and then to SNA
via the  Heagy-Hammel(HH) mechanism.  This transition occurs in the
range 0.953 $< a <$ 0.958 and 0.38 $< E_1 <$ 0.58.  It is denoted as {\bf HH}
in Fig. 2.

Moving higher up in the amplitude space and  from  right to  left, we find that SNAs and
eventually chaos occur from period three-quasiperiodic attractor via type I
intermittency route as $a$ is varied in the narrow range   0.949 $< a <$ 0.954 
and for $E_1$ in the range 0.623 $<E_1<$ 0.645.  It is denoted as  {\bf INT}
within a small box [Fig. 2(a)].  In Fig. 2(b), the enlarged portion of
the box in Fig. 2(a) shows the region of existence of the intermittent SNA
occuring between quasiperiodic and chaotic attractors.
  
In this section, we have identified atleast three interesting dynamical
features namely, (1) Heagy-Hammel, (2) fractalization and (3) type I
intermittent routes in the two-parameter diagram.  Now, we describe each one of
the them in detail from the point of view of numrical analysis as well as 
experimental identification as follows.

\section{Heagy-Hammel route to SNA}
The first of these routes that we encounter is the Heagy-Hammel route in which a
period$-2^k$ torus gets wrinkled and upon collision with its unstable parent
period$-2^{k-1}$ torus bifurcates into a SNA.  Such a behavior has been observed in the present quasiperiodically  forced
negative conductance series LCR circuit within the range of  $a$  values,
0.953$<a<$ 0.958, and $E_1$ values, 0.38$<E_1<$ 0.58, while the other parameters 
are fixed as prescribed earlier in section II.  

\subsection{Numerical Analysis}
\begin{figure}
\centering
\includegraphics[width=1.0\columnwidth]{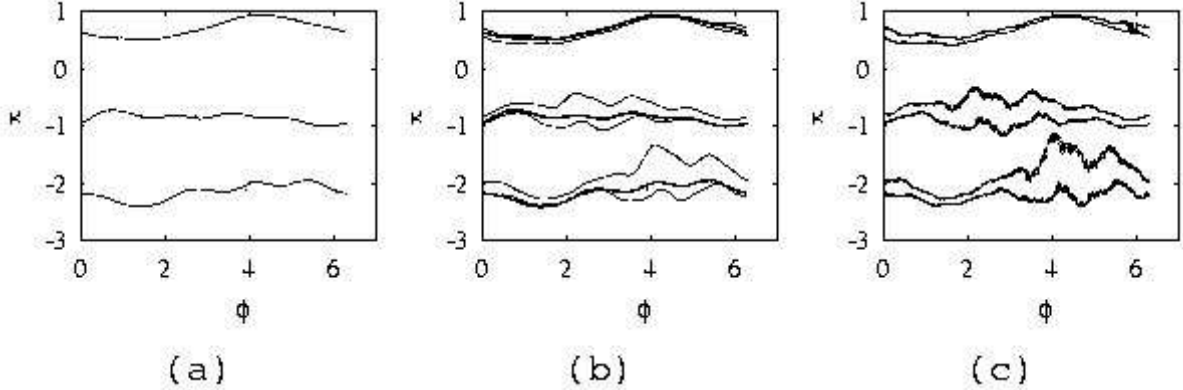}
\caption{\label{hhqp}Projection of the numerically simulated attractors of Eqs. (2) in the $(\phi - x)$ plane
for fixed $E_1 =$0.44  and various values of $a$ indicating the transition from
quasiperiodic attractor to SNA through Heagy-Hammel mechanism: (a) period-3
torus (3T) for $a$=0.95632, (b) period-6 torus (6T) for $a$=0.95593, and 
(c) SNA at $a$=0.95592.}
\end{figure}
     
More specifically, let us fix the parameter $E_1$ at $E_1=$ 0.44,  while
decreasing the value of $a$.  For $a=$ 0.95632, the circuit equation (2)
associated  with Fig. 1 is found to exhibit a period-3 torus attractor denoted
as 3T (see Fig. 2)  and the  Poincar\'e map has three smooth branches (Fig.
3(a)), whose phase portrait and power spectrum are shown in Figs.~4a(i) and
4a(ii).  As the value of $a$ is decreased to $a=$ 0.95593, the attractor
undergoes torus doubling bifurcation and the corresponding period-6
quasiperiodic orbit is denoted as 6T in Fig. 2 and the  Poincar\'e map has six
smooth branches as seen in Fig.~3(b).  The corresponding phase portrait and
power spectrum are shown in Figs.~4b(i) and 4b(ii).  In the generic case, the
period-doubling occurs  in an infinite sequence until the accumulation point is
reached, beyond which  chaotic behavior appears. However, with tori, in the
present case, further  torus doubling does not takes place, but the torus
becomes wrinkled; that  is the truncation of the three torus doubling begins
when the six strands  of the 6T attractor become extremely wrinkled.   This is
because the period-doubled six torus collides with its unstable parent, and
this occurs only for a few narrow selected parameter intervals, when $a$ is
decreased to $a=$ 0.95593 as shown in Fig.3(b).  For example, when the value of
$a$ is decreased to $a=$ 0.95592, the attractor becomes extremely wrinkled. 
During this transition, the strands are seen to come closer to the unstable
period 3T orbit and lose their continuities when the strands of torus doubled
orbit collide with unstable parent and ultimately result in a fractal
phenomenon as shown in Fig. 3(c) when $a$ is decreased to $a=$ 0.95592.  The
phase portrait and power spectrum corresponding to Fig.~3(c) are shown in
Figs.~4c(i) and 4c(ii).  At such a value, the attractor, Fig. 3(c), possesses a
geometrically strange property but does not exhibit sensitivity to initial
conditions [the maximal Lyapunov exponent is negative as seen in Fig. 5(a)] and
so it is indeed a strange nonchaotic attractor. As $a$ is decreased further to
$a=$ 0.95435, the attractor has eventually a positive Lyapunov exponent and
hence it corresponds to chaotic attractor (denoted C in Fig. 2).  
 
Now we examine the Lyapunov exponent for the transition from period-3 torus
doubling to SNA. During this transition, the largest maximal Lyapunov exponent
\begin{figure}
\centering
\includegraphics[width=0.9\columnwidth]{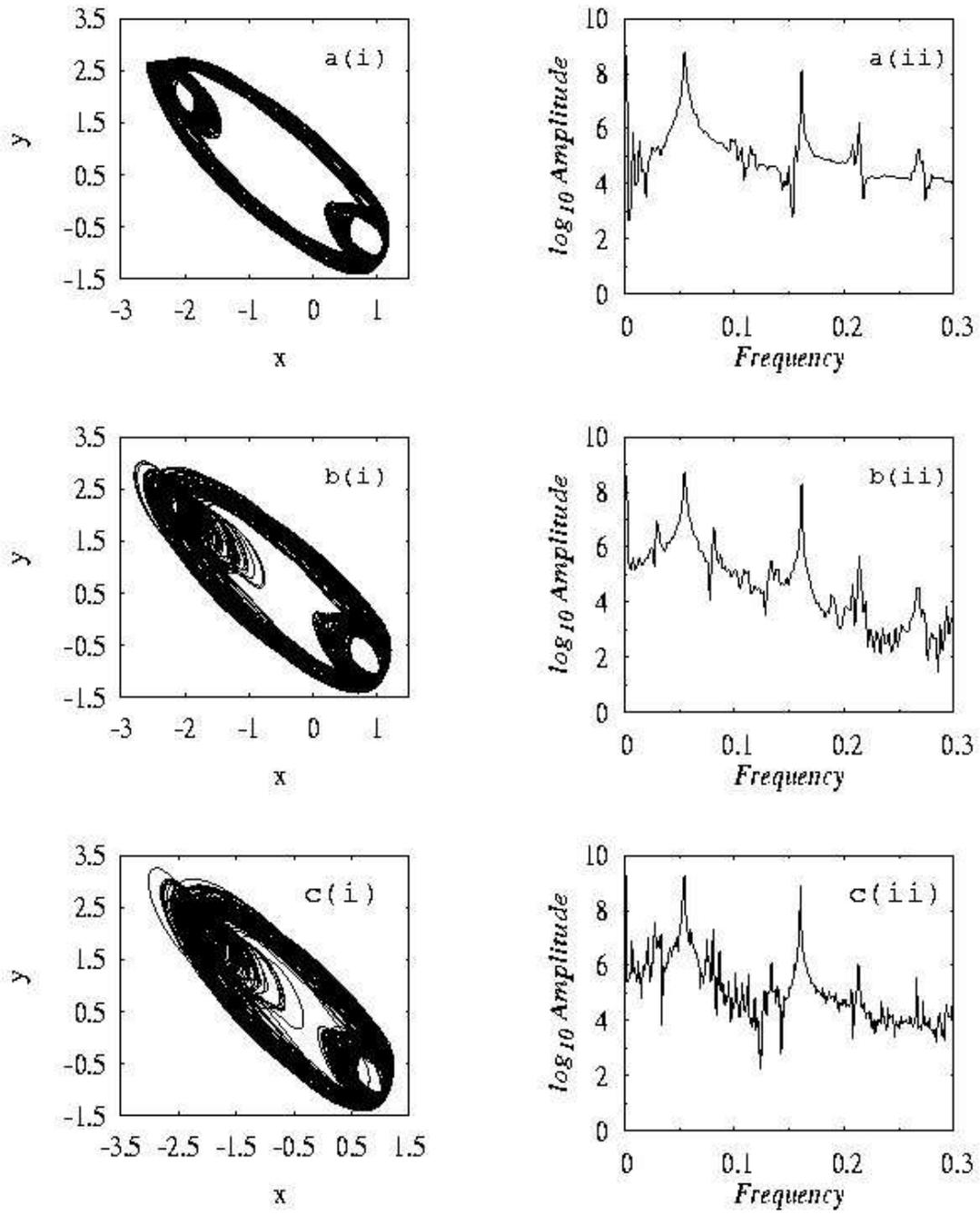}
\caption{\label{hhnum}Projection of the numerically simulated attractors of 
Eqs. (2) in the $(x, y)$ plane
for fixed $E_1 =$0.44  and various values of $a$ indicating the transition from
quasiperiodic attractor to SNA through Heagy-Hammel route: (a) period-3
torus (3T)
for $a$=0.95632, (b) period-6 torus (6T) for $a$=0.95593 and (c) SNA at $a$=0.95592:
(i) phase trajectory in the $(x-y)$ space; (ii) power spectrum.}
\end{figure}
$\Lambda$ as a function of $a$ for a fixed $E_1=$ 0.44  remains negative,
which is shown in Fig.~5(a). Hence, the attractor is strange but nonchaotic. 
We also note that there is an abrupt change in the maximal Lyapunov exponent
during the transition from period-3 torus doubled attractor to SNA and its
variance (Fig. 5(a) \& 5(b)).   When we examine this in a sufficiently small
neighborhood of the critical value $a_{HH}=$0.95593, the transition is clearly
revealed by the Lyapunov exponent which varies smoothly in the torus region 
($a < a_{HH}$) while it varies irregularly in the SNA region  ($a > a_{HH}$).
It is also possible to identify this transition point by examining the variance
of Lyapunov exponent, as shown in Fig. 5(b) in which the fluctuation is small
in the torus region while it is large in the SNA region.
\begin{figure}
\centering
\includegraphics[width=0.95\columnwidth]{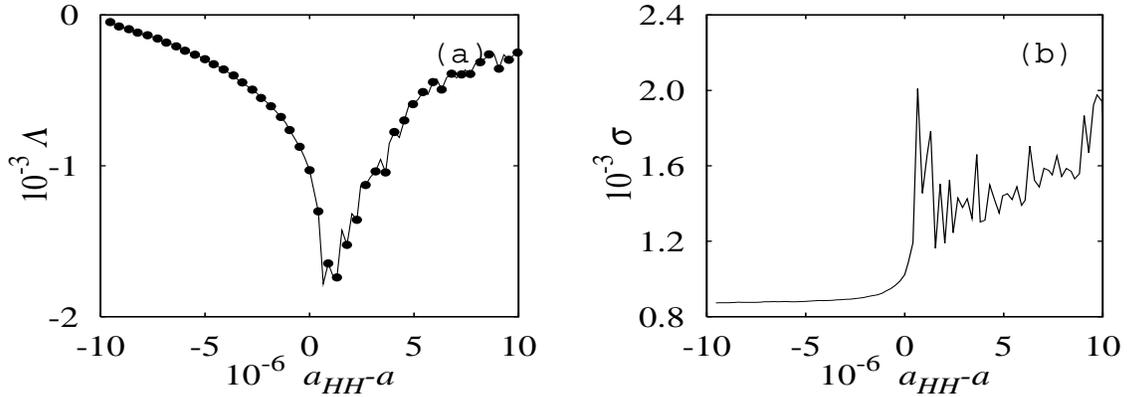}
\caption{\label{hhlya}Transition from three doubled torus to SNA through
Heagy-Hammel route in region HH obtained from numerical data:  (a) the behavior of the
maximal Lyapunov exponent $( \Lambda)$ and (b) the variance $( \sigma)$ for
$E_1=$0.44.}
\end{figure}
\subsection{Experimental Confirmation}
To confirm that the above results hold good in the actual experimental circuits (Fig.~1) also,  the phase trajectory is obtained  experimentally by
measuring the voltage $v$ across the capacitor $C$ and the current $i_L$
through the inductor $L$  in the circuit (Fig.~1)  and connecting them to the $X$ and $Y$
channels of an oscilloscope.  Then, a live picture of the corresponding power
spectrum (obtained from a digital storage oscilloscope - HP 54600 series) of
the projected attractor has also been used to distinguish the different
attractors. The experimentally measured phase portraits and Fourier spectra
shown in Figs.~6 correspond to the transition from period-3 torus
quasiperiodic attractor to SNA  through the HH mechanism as shown in Figs.~3 and 4.
It has been found that the simulated results and experimentally observed
results in the phase-space as well as power spectra are qualitatively similar to each other. In
particular, in both cases, the spectra of the quasiperiodic attractors
are concentrated at a small discrete set  of frequencies while the spectra of SNA have
a much richer harmonic. 
\begin{figure}
\centering
\includegraphics[width=0.5\columnwidth]{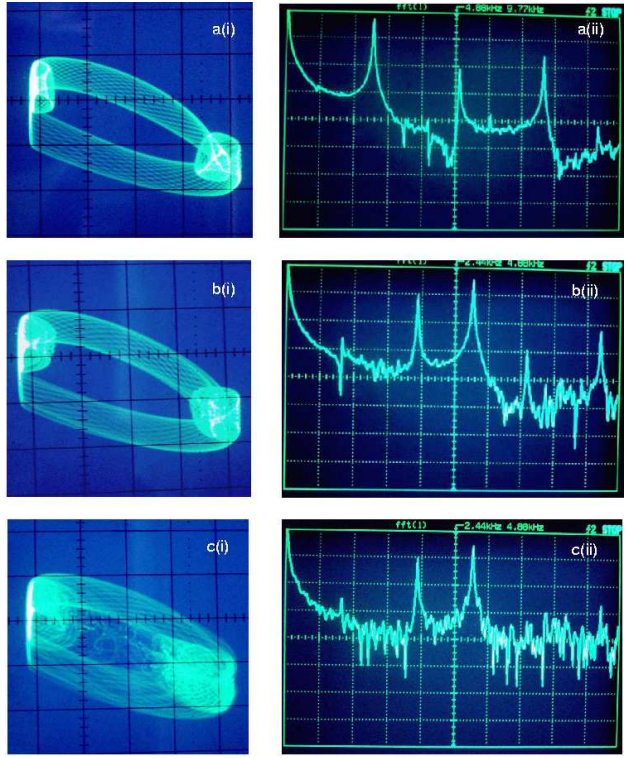}
\caption{\label{hhphase}(Color online) Attractors obtained experimentally from the circuit
given in Fig. 1 corresponding to Figs. 4. (a) period-3 torus (3T) for R=2109
$\Omega$, (b) period-6 torus (6T) for R=2106 $\Omega$ and (c) SNA at R=2104
$\Omega$ for fixed value of $E_{f1}$=0.22 V: (i) phase trajectory in the 
$(v -i_L)$ space;
(ii) power spectrum.}
\end{figure}
To distinguish further in the characteristic aspect that the attractors
depicted in Figs. 3, 4 \& 6 are  quasiperiodic and strange nonchaotic, we
proceed to  quantify the changes in the power
\begin{figure}
\centering
\includegraphics[width=0.9\columnwidth]{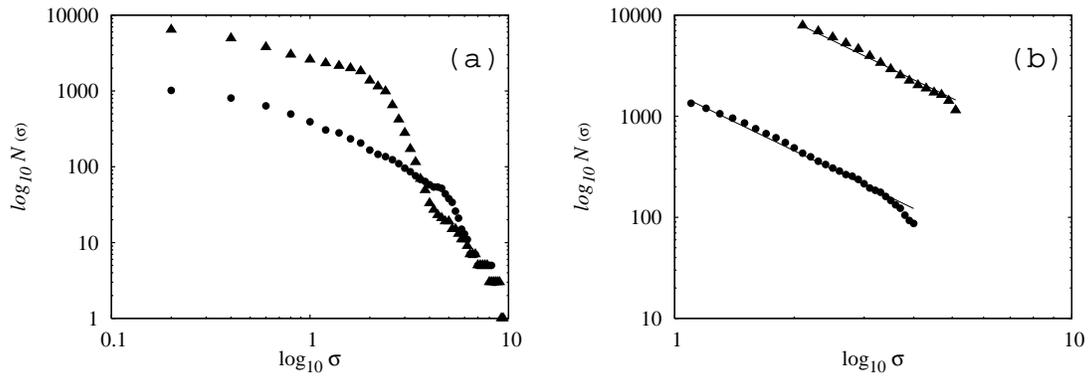}
\caption{\label{hhspecdst}Spectral distribution function for the quasiperiodic 
attractors and SNAs created through
the Heagy-Hammel route: (a) quasiperiodic attractor, (b) strange nonchaotic attractor. Here 
the numerical study is indicated by filled circles, and experimental
result is denoted by filled triangles.}
\end{figure}
spectrum.  The spectral distribution (which is defined as the number of peaks
in the  Fourier amplitude spectrum larger than some value say $\sigma$) for
quasiperiodic attractor and SNA are shown in Figs~7. In Figs.~7 the filled circles and
the filled triangles  denote the spectral distribution obtained through
numerical simulation and  experimental measurements respectively.  The
experimental data are recorded using a 16-bit data acquisition system
(AD12-16U(PCI)EH) at the sampling rate of 200 kHz. It is found numerically as
well as experimentally that the quasiperiodic attractors obey a scaling
relationship $N(\sigma)$=$\log_{10}(1/ \sigma)$ [see Fig.~7(a)] while the SNAs
satisfy a scaling power law relationship $N(\sigma)=$ $\sigma^{-\beta}$,
$1<\beta<2$. The approximate straight line in the log-log plot shown in Fig.
7(b) obeys the power-law relationship with a value of $\beta=$ 1.9 for
numerical study and 1.84 for experimental study. 

It has also been found that a typical trajectory on a SNA actually possesses
positive Lyapunov exponents in finite time intervals, although the asymptotic
exponent is negative.  As a consequence, one observes the different
characteristics of SNA created through different mechanisms by a study of the
differences in the distribution of finite-time exponents $P(N ,\lambda)$
~\cite{apvm1997}.  For each of the cases, the distribution can be obtained by
taking a long trajectory and dividing it into segments of length $N$, from
which the local Lyapunov exponent can be calculated.  In the limit of large
\begin{figure}
\centering
\includegraphics[width=1.0\columnwidth]{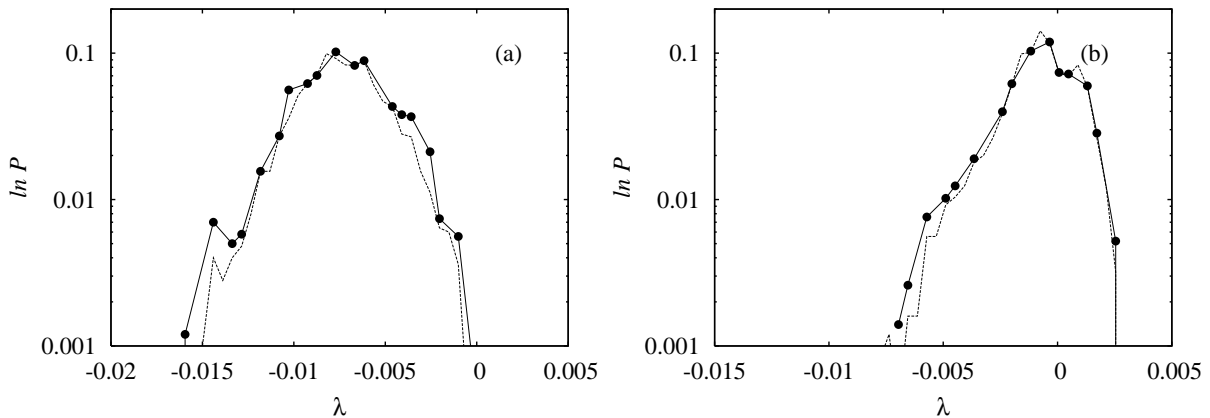}
\caption{\label{ftlya_hh}Distribution of finite-time Lyapunov exponents on SNAs
created through  the Heagy-Hammel route: (a) quasiperiodic attractor, 
(b) strange nonchaotic attractor. Finite-time Lyapunov exponents calculated from
numerical data are indicated by dashed lines, and from experimental
data are denoted by solid lines.}
\end{figure}
$N$, this distribution will collapse to a $\delta$ function  $P(N,\lambda)
\longrightarrow (\delta - \lambda)$.  The deviations from and the approach to
the limit can be very different for SNAs created through different mechanisms.
We apply Wolf algorithm to determine the Lyapunov exponents from the
experimental data~\cite{awjbs1985}. 
Fig.~8 illustrates the distributions for $P(2000, \lambda)$ which is strongly
peaked about the Lyapunov exponent when the attractor is a torus, but on the
SNA the distribution picks up a tail which extends into the local Lyapunov
exponent $\lambda > $0 region. (Finite-time Lyapunov exponents calculated from
numerical data are indicated by dashed lines, and from experimental
data are denoted by solid lines) This tail is directly correlated with enhanced
fluctuation in the Lyapunov exponent on SNAs.  On Heagy-Hammel SNA, the 
distribution shifts continuously to larger Lyapunov exponents. 
Further the shapes
for the torus regions [Fig.~8(a)] and SNA regions [Fig.~8(b)] are very
different. The results clearly confirm
that the HH mechanism is operative in the parameter regime of the present discussion. 
 
\section{Fractalization route to SNA}
The second one of the routes we have identified in the present system is the gradual
fractalization route where a torus gets increasingly wrinkled and then transits
to a SNA without interaction (in contrast to the previous case of HH) with a
nearby unstable orbit as we change the system parameter.   In this route  a
period-3$^k$ torus becomes wrinkled and then the wrinkled attractor gradually
loses its smoothness and forms a 3$^k$-band SNA as we change the system
parameter $a$ for fixed value of $E_1$.
\begin{figure}
\centering
\includegraphics[width=1.0\columnwidth]{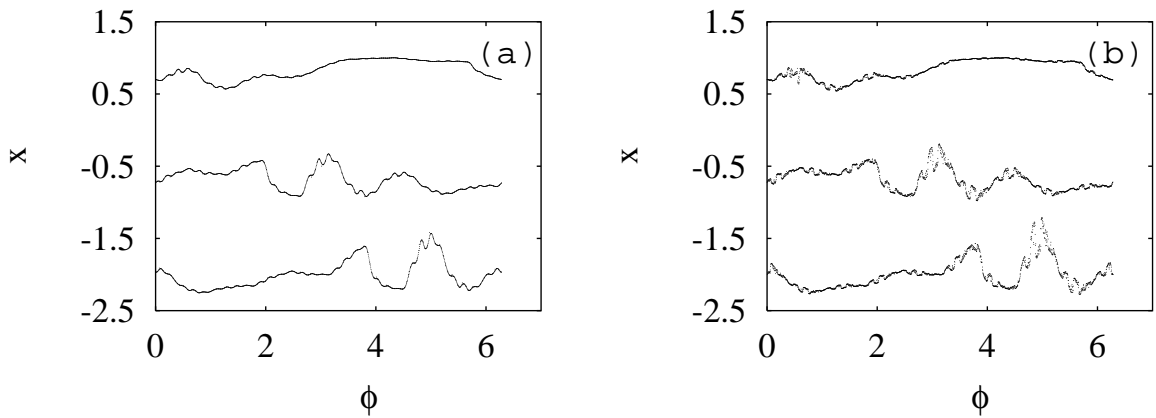}
\caption{\label{fqp}Projection of the numerically simulated  attractors of Eqs.~(2) in the $(\phi - x)$ plane
for fixed $E_1 =$0.34  and various values of $a$ indicating the transition from
quasiperiodic attractor to SNA through fractalization route. (a) period-3 torus (3T)
for $a$=0.954406 and (b) SNA at $a$=0.954351.}
\end{figure}
The qualitative (geometric) structure of the attractor remains more or less
the same during the process. Such a phenomenon has been observed in the present
circuit in two different regions indicated as F  in Fig.~2 for certain ranges of
$a$ in the regions of interest. 

\subsection{Numerical Analysis} 
Now let us consider  the phase diagram (Fig.~2)
where we have identified such type of fractalization. To exemplify the nature
of this transition, we fix the parameter $E_1$ at $E_1=$0.34, and vary $a$ in
the range  0.953$<a<$ 0.955 (Fig.~2).   On decreasing the $a$ value,
oscillations of torus (3T) in the amplitude direction starts  to appear at
$a=0.954406$ (Fig.~9(a)) whose phase portrait and power spectrum are shown in
Figs.~10a(i) and 10a(ii). As $a$ is decreased further to $a=0.954351$, the
oscillatory behavior  of the torus  gradually approaches a fractal nature.  The
torus (3T) attractor  
\begin{figure}
\centering
\includegraphics[width=0.8\columnwidth]{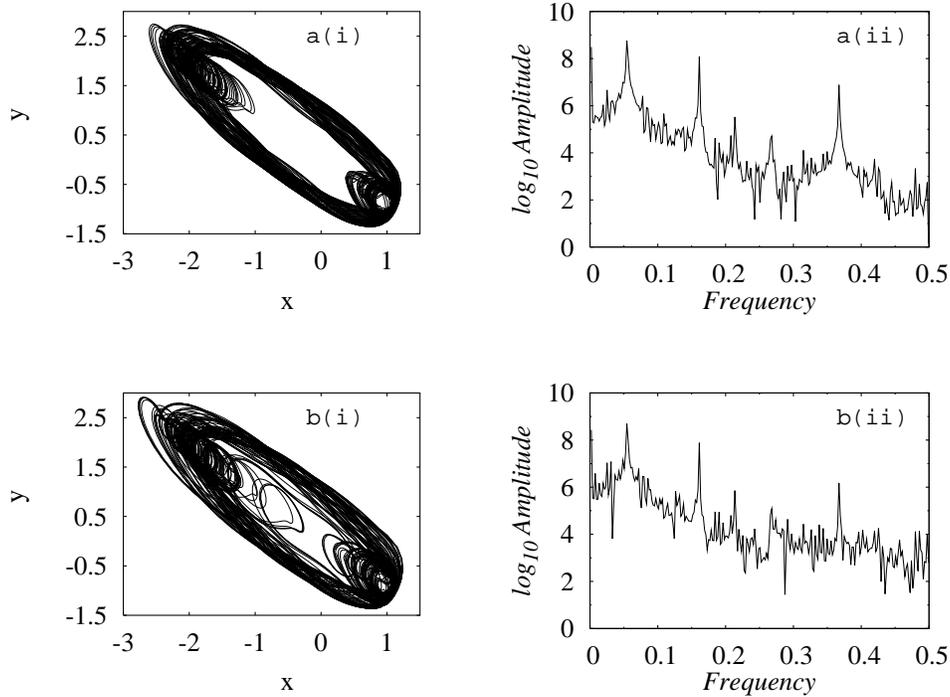}
\caption{\label{fnumphas}Projection of the numerically simulated attractors of Eqs. (2) in the $(x,
y)$ plane for fixed $E_1 =$0.34 and various values of $a$ indicating the
transition from quasiperiodic attractor to SNA through fractalization
route:(a) period-3 torus (3T) for $a$=0.954406 and (b) SNA at $a$=0.954351:
(i) phase trajectory in the $(x-y)$ plane ; (ii) power spectrum.}
\end{figure}
gets increasingly wrinkled and transforms into a SNA  at $a_{GF}=0.954351$ as shown
in Fig. 9(b).  The corresponding phase portrait and power spectrum are shown in
Fig.~10b(i) and 10b(ii).

At such values,  the nature of the attractor is strange (see Fig. 9(b))
eventhough the largest Lyapunov exponent in Fig. 11(a) remains negative. It is
very obvious from these transitions that the 3 torus with three smooth branches
in the  Poincar\'e map (Fig. 9(a)) gradually losses its smoothness and
ultimately approaches a fractal behavior via a SNA (in Fig. 9(b)) before the
onset of chaos as the parameter $a$ decreases further.  Such a phenomenon is
essentially a gradual fractalization of the  torus as was shown by Nishikawa
and Kaneko~\cite{tnkk1996} in their route to chaos via SNA. In this route,
there is no collision involved among the orbits and therefore the Lyapunov
exponent and its variance change only slowly as shown in Fig.~11(a) and 11(b)
and there are no significant changes in its variance (see Fig. 10(b)).  At even
lower values of `$a$',  $a=$ 0.954, the circuit exhibits chaotic oscillations
as shown in region C of Fig. 2.

\subsection{Experimental confirmation}
To confirm the numerical results further,  experimentally measured phase
portraits and Fourier spectrum results corresponding to the circuit
of Fig.~1 are presented in Figs.~12 which correspond to the
transition from quasiperiodic attractor to SNA  through gradual fractalization
shown in Figs.~9 and 10. It has been noticed that the simulated results and
experimentally measured results in the phase-space as well as power spectrum
are in close agreement.
\begin{figure}
\centering
\includegraphics[width=0.9\columnwidth]{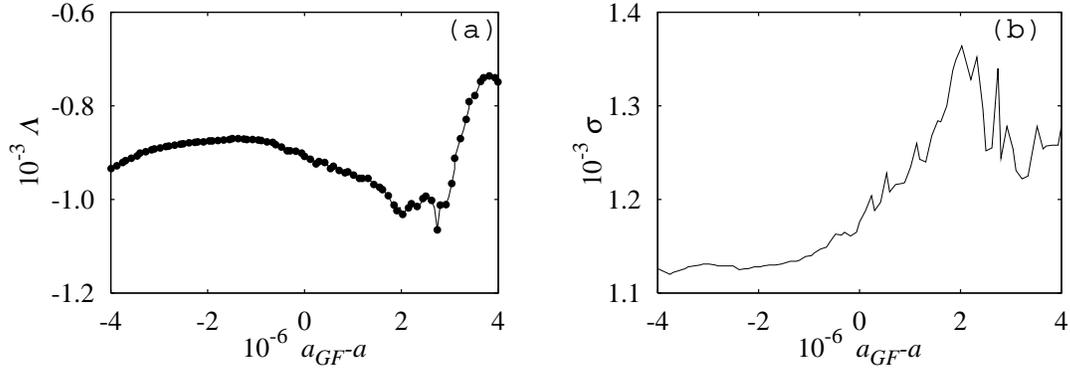}
\caption{\label{flya}Transition from three torus to SNA through fractalization 
route obtained from numerical data:  (a) the behavior of the maximal Lyapunov exponent $( \Lambda )$
and (b) the variance $( \sigma)$ for $E_1$=0.34.}
\end{figure}
\begin{figure}
\centering
\includegraphics[width=0.5\columnwidth]{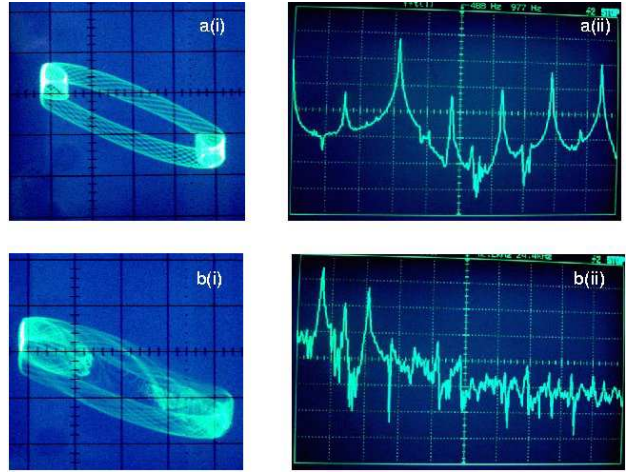}
\caption{\label{fexpphase}(Color online) Attractors obtained experimentally from the circuit
given in Fig.1 corresponding to Figs. 10. (a) period-3 torus (3T) for R=2102
$\Omega$ and (b) SNA at R=2101 $\Omega$ for fixed value of $E_{f1}$=0.17 V: (i)
phase trajectory $(v -i_L)$; (ii) power spectrum.}
\end{figure} 
To verify further whether the attractors depicted in Figs.~10 and 12 are
quasiperiodic and strange nonchaotic attractors, we proceed to quantify the
changes in the numerically and experimentally measured power spectra. In our
analysis it has been verified that the quasiperiodic attractor obeys a scaling
relationship  $N(\sigma)$=$\log_{10}(1/ \sigma)$[see Fig.13(a)] while in the
approximate straight line shown in the log-log plot Fig. ~13(b) satisfying  the
power relationship $N(\sigma)=$ $\sigma^{-\beta}$, with an estimated value of
$\beta$=1.78 for simulation and $\beta$=1.9 for experimental measurement
confirms that the attractor created through this mechanism is indeed a strange
nonchaotic attractor.

Fig.~14  illustrates the distributions for $P(2000, \lambda)$ which is strongly
peaked about the Lyapunov exponent when the attractor is a torus, but on the
SNA the distribution picks up a tail which extends into the local Lyapunov
exponent $\lambda > $0 region.  This tail is directly correlated with enhanced
fluctuation in the Lyapunov exponent on SNAs.  On the fractalized SNA , the
distribution shifts continuously to larger Lyapunov exponents, but the shape
remains the same for torus regions as well as SNA regions.

\begin{figure}
\centering
\includegraphics[width=0.7\columnwidth]{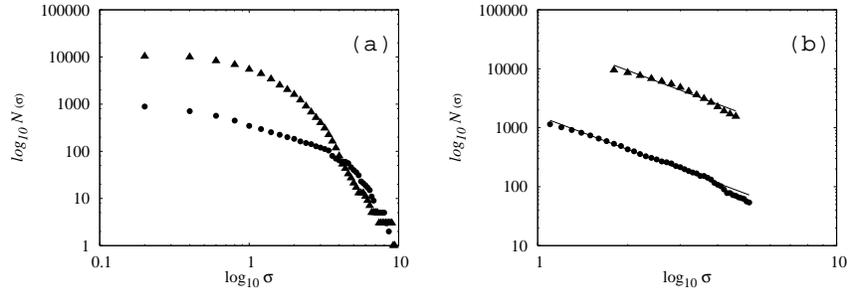}
\caption{\label{fspecdst}Spectral distribution function for spectra of quasiperiodic attractor and
SNAs created through
gradual fractalization route: (a) quasiperiodic attractor, (b) strange nonchaotic attractor.Here
numerical study is indicated by the filled circles and experimental study is denoted by the
filled triangles.}
\end{figure}  
\begin{figure}
\centering
\includegraphics[width=1.0\columnwidth]{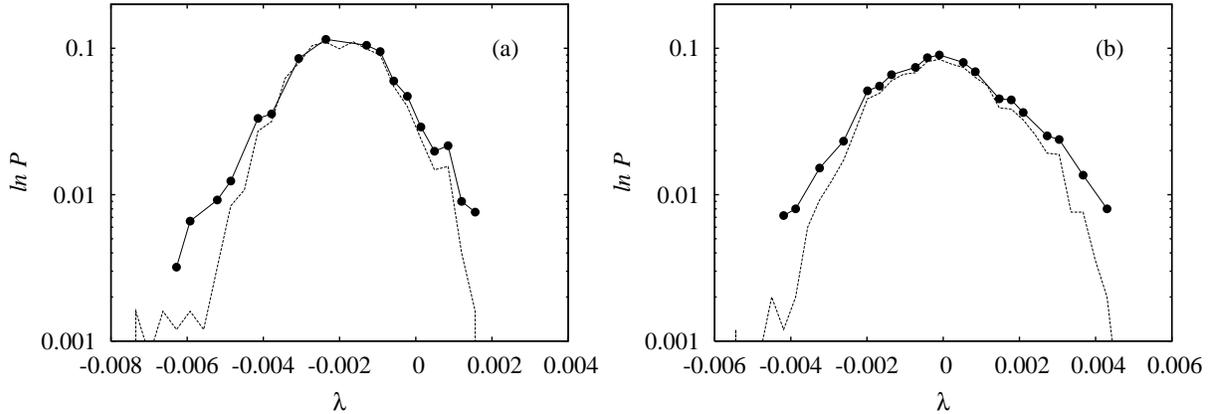}
\caption{\label{ftlya_fra}Distribution of finite-time Lyapunov exponents on SNAs
created through gradual fractalization.: (a) quasiperiodic attractor, 
(b) strange nonchaotic attractor. Finite-time Lyapunov exponents calculated from
numerical data are indicated by dashed lines, and from experimental
data are denoted by solid lines.}
\end{figure}
\section{Intermittent route to SNA}
Finally, the  third of the  routes that is predominant in this system is an
intermittent route in which  the torus is eventually replaced by a   strange
nonchaotic attractor  through an analog of the saddle-node bifurcation.    
\begin{figure}
\centering
\includegraphics[width=0.9\columnwidth]{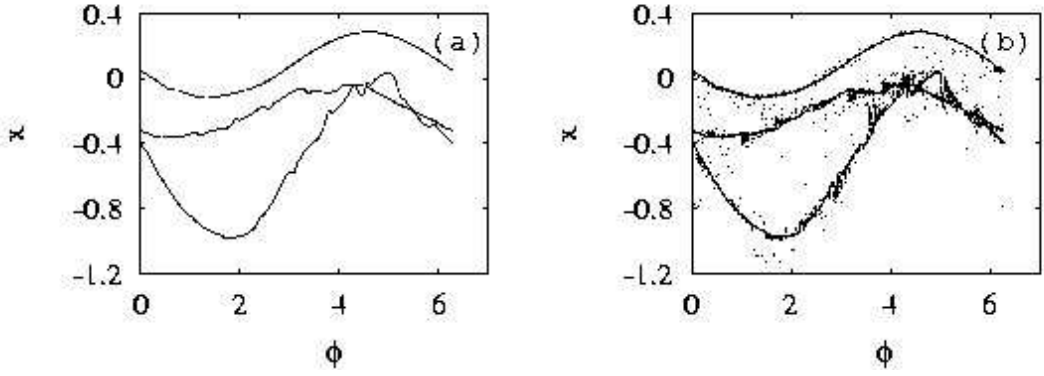}
\caption{\label{iqp}Projection of the simulated attractors of Eqs.~(2) in the $(\phi-x)$
plane for fixed $E_1 =$0.635  and various value of a: indicating the transition
from quasiperiodic attractor to SNA through type I intermittent route. (a)
torus (3T) for a=0.951912 and (b) SNA at a=0.951889.}
\end{figure}
Such a phenomenon has been identified within the range $0.623 < E_1 < 0.645$
for the amplitude while  the parameter $a$ is decreasing from right to left in
the narrow range 0.949 $< a <$ 0.954 for fixed $E_1$.
\begin{figure}
\centering
\includegraphics[width=0.85\columnwidth]{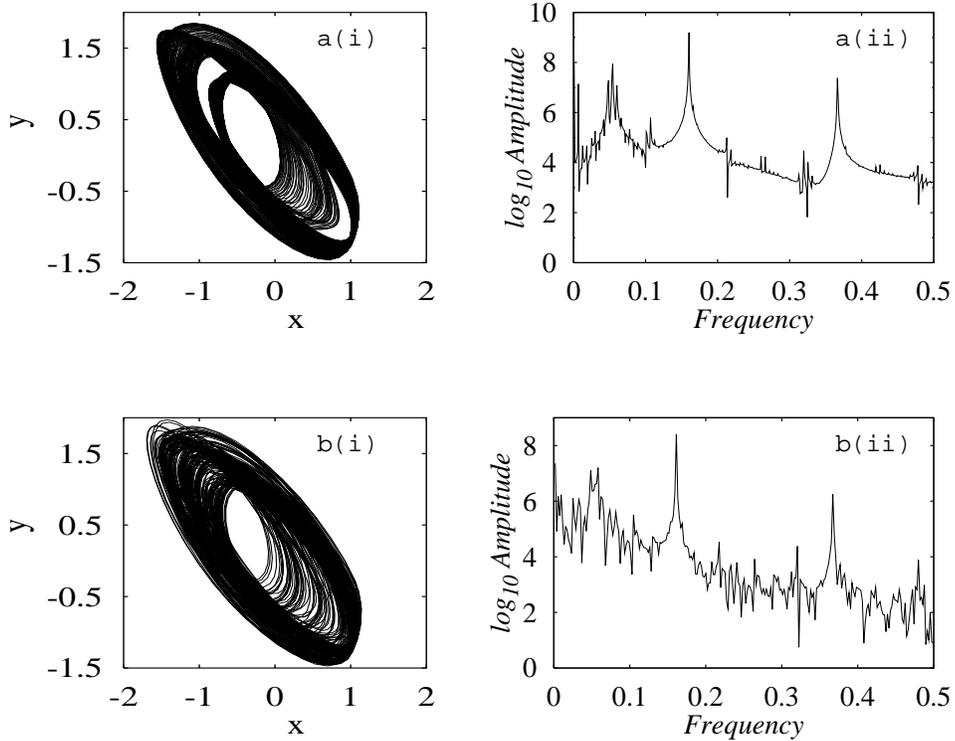}
\caption{\label{inumphas}Projection of the numerically simulated attractors of Eqs. (2) for fixed $E_1 =$0.635 and two
different values of $a$ indicating the
transition from quasiperiodic attractor to SNA through type I intermittent
route: (a) period-3 torus (3T) for $a$=0.951912 and (b) SNA at $a$=0.951889:
(i) phase trajectory in the $(x-y)$ plane; (ii) power spectrum.}
\end{figure}
\subsection{Numerical Analysis} 
To illustrate the above transition,  let us fix the parameter $E_1$ at $E_1=
0.635$ while  $a$ is decreased from $a =0.95192$.  Figure 15(a) shows the
projection of  a three-period quasiperiodic attractor which has three smooth
branches in the Poincar\'e section. The corresponding phase portrait and power
spectrum are shown in Figs.~16a(i) and 16a(ii). As $a$ is decreased further,
the attractor starts to wrinkle.  On further decrease of $a = 0.951889$,  the
attractor becomes extremely wrinkled and has several sharp bends.  
\begin{figure}
\centering
\includegraphics[width=0.9\columnwidth]{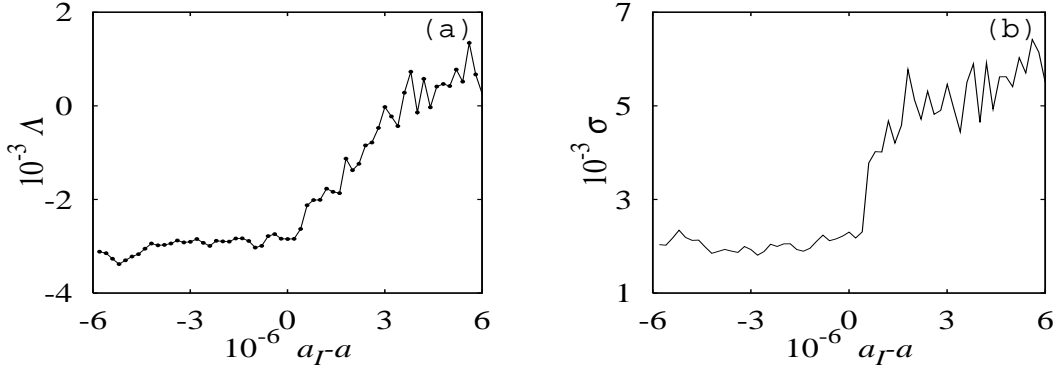}
\caption{\label{ilya}Transition from three torus to SNA through type I
intermittent  mechanism obtained from numrical data:  (a) the behavior of the maximal Lyapunov
exponent $( \sigma)$ and (b) the variance $( \Lambda)$ for $E_1 =0.635$.}
\end{figure}
However, as `$a$` passes a threshold value $a_I=0.951876$, an intermittent 
transition from the torus to SNA occurs. At the intermittent transition, the
amplitude variation loses its regularity and a burst appears in the regular
phase (quasiperiodic orbit trajectory). The duration of laminar phases in this
state is random. An example of the transition to such SNAs is shown in
Fig.~15(b), the corresponding phase portrait and power spectrum are shown in
Figs.~16b(i) and 16b(ii). At this transition,  we also note that there is an
abrupt change in the maximal Lyapunov exponent as well as its variance
corresponding to the characteristic signature of the intermittent route
[indicated in Figs. 17(a) and 17(b)] to SNA. 

In the HH case, the points on the SNA are distributed over the entire region
enclosed by the wrinkled bounding torus, while in the fractalization case the
points on the SNA are distributed mainly on the boundary of the torus. 
Interestingly, in the present case shown in Fig. 15(b), most of the points of
the SNA remain within the wrinkled torus with sporadic large deviations.  The
dynamics at this transition obviously involves a kind of intermittency.  Such
an intermittency transition could be characterized by scaling behavior.  The
laminar phase in this case is the torus while the burst phase is the nonchaotic
attractor.  In order to calculate the associated scaling constant, we coevolve
the trajectories for two different values of $a$, namely, $a_c$ and another
value of a near to $a_c$, while keeping identical initial conditions $(x_i,
\theta_i)$ and the  same parameter value $E_1$.  As the angular
coordinate $ \theta_i$ remains identical, the difference in $x_i$ allows one to
compute the average laminar length between the bursts.  The plot of average
laminar length $<l>$ for this attractor reveals a power law relationship of the
form 
\begin{eqnarray}
<l> = ( a_{\text{critical}} - a)^{- \alpha}.
\end{eqnarray}
with the estimated value of $\alpha=$0.31 (see Fig.~18). This analysis also
confirms that such an attractor is associated with standard intermittent
dynamics of type I described in Ref.~\cite{yppm1980,mlsr2003,eo1994}. 
\begin{figure}
\centering
\includegraphics[width=0.7\columnwidth]{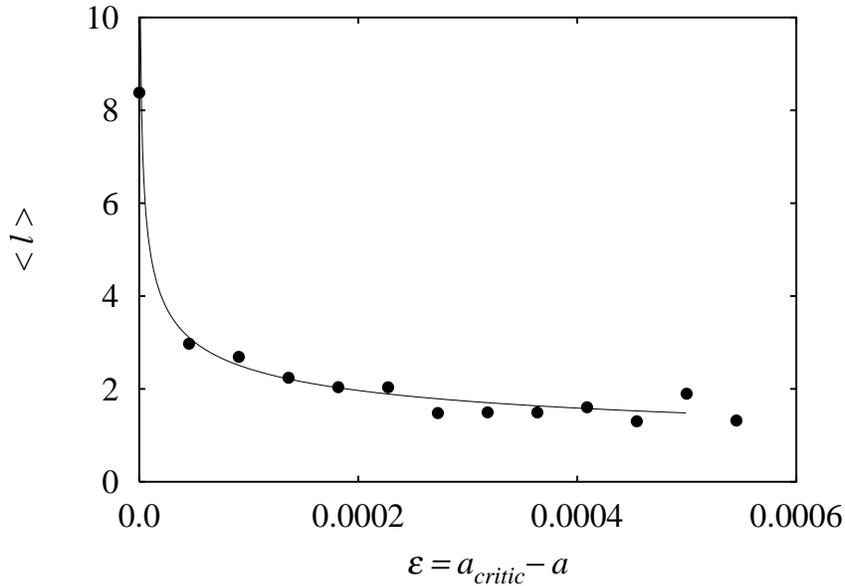}
\caption{\label{sclaw}Average laminar length $(<l>)$ vs $(a_{critic}-a)$ at
$a_{critic}=0.951876$ obtained from numerical data.}
\end{figure}
 
\subsection{Experimental Confirmation}
Next, we compare the simulation results in Figs.~16 and the experimental 
results given in Figs.~19. The range of parameters  chosen  for experimentally
measured phase portraits and Fourier spectra results given in Figs.~19 
correspond to the transition from quasiperiodic attractor to SNA  through
intermittent nature shown in Figs.~15 and 16. It has been found that the
simulated results and experimentally observed results in the phase-space as
well as power spectrum appears to be qualitatively similar in nature.  To
distinguish  further that the attractors depicted in  Figs.~15, 16 and 19 are
quasiperiodic and  strange nonchaotic attractors, the numerically and
experimentally measured  power spectra are quantified.  It has been noted that
the quasiperiodic attractor obeys a scaling relationship 
\begin{figure}
\centering
\includegraphics[width=0.5\columnwidth]{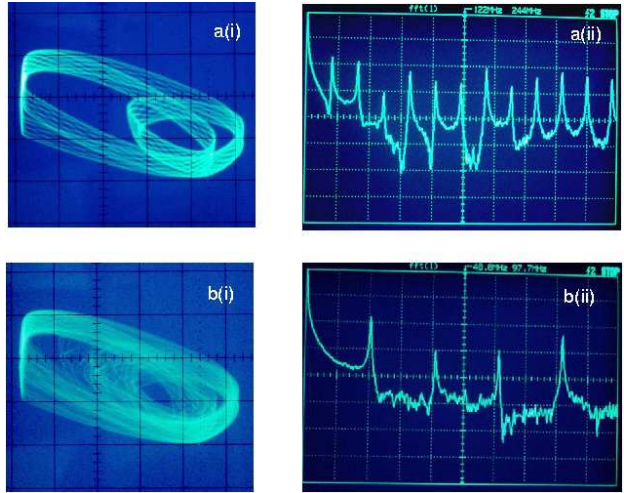}
\caption{\label{iexpphas}(Color online) Attractors obtained experimentally from the
circuit given in Fig.1 corresponding to Figs.~\ref{inumphas}.
(a) period-3 torus (3T) for R=2099
$\Omega$ and (b) SNA at R=2097 $\Omega$ for fixed value of $E_{f1}$=0.318 V: (i)
phase trajectory $(v -i_L)$; (ii) power spectrum.}
\end{figure}
$N(\sigma)$=$\log_{10}(1/ \sigma)$ [see Fig.~20(a)] while the SNAs created through
this mechanism satisfy a scaling power law relationship $N(\sigma)=$
$\sigma^{-\beta}$, $1<\beta<2$. The approximate straight line in the log-log
plot shown in Fig. 20(b) obeys the power-law relationship with a value of
$\beta = $ 1.86  and 1.89 for numerical simulation and experimental measured 
studies respectively. 
\begin{figure}
\centering
\includegraphics[width=0.9\columnwidth]{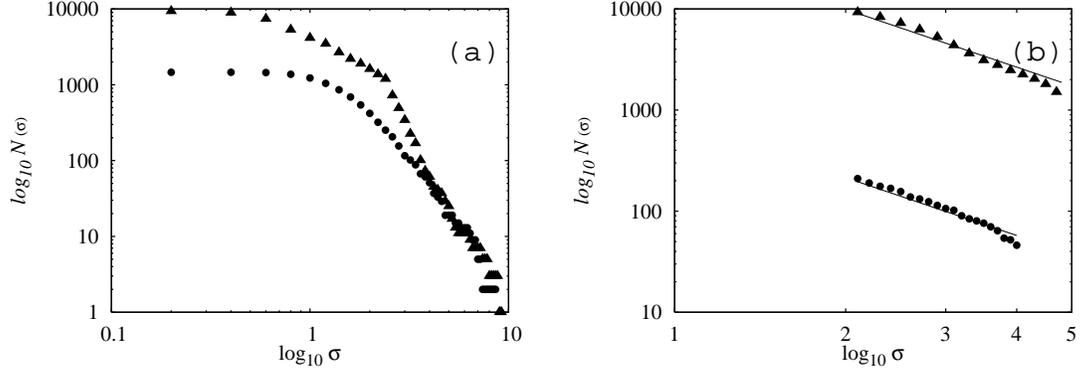}
\caption{\label{ispecdst}Spectral distribution function for spectrum SNAs
created through type-I intermittency route (circle denotes numerical study 
and triangle indicates experimental 
study).}
\end{figure}
Fig.~ 21 illustrates the distributions for $P(2000, \lambda)$ which is strongly
peaked about the Lyapunov exponent when the attractor is a torus, but on the
SNA the distribution picks up a tail which extends into the local Lyapunov
exponent $\lambda > $0 region.  This tail is directly correlated with enhanced
fluctuation in the Lyapunov exponent on SNAs.   On the intermittent SNA route,
the actual shapes of distribution on the torus and the SNA are very different. 
\begin{figure}
\centering
\includegraphics[width=1.0\columnwidth]{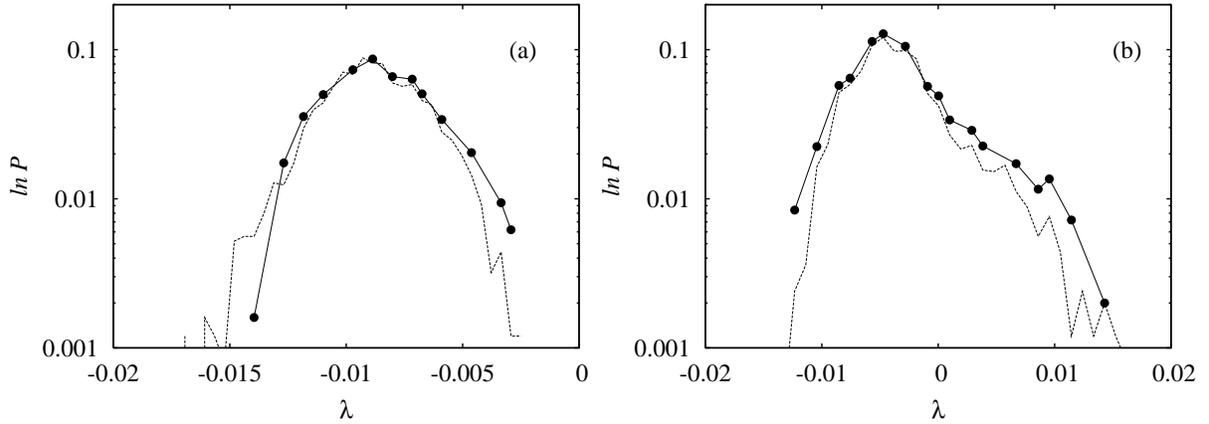}
\caption{\label{ftlya_int}Distribution of finite-time Lyapunov exponents on
SNAs created through  type-I intermittency.  (a) quasiperiodic attractor,  (b)
strange nonchaotic attractor. Finite-time Lyapunov exponents calculated from
numerical data are indicated by dashed lines, and from experimental data are
denoted by solid lines.}
\end{figure}

\section{Summary and conclusion} In this paper,  various transitions from
the quasiperiodic attractors to  the strange nonchaotic attractors are demonstrated
experimentally in a simple quasiperiodically  driven electronic system.
Specifically, the three prominent routes, namely Heagy-Hammel, fractalization
and type I intermittent routes for the creation of SNAs are demarcated the
different regions in the ($a-E_1$) parameter space.   First, we have used simulation
results to show the bifurcation process of this circuit from the quasiperiodic
attractors to the strange nonchaotic attractors.  Then we have experimentally observed
the existence of the strange nonchaotic attractors as a part of the whole
bifurcation process as predicted by the simulation.  The experimental
observations,  numerical simulations and characteristic analysis show that the
simple dissipative quasiperiodically forced negative conductance series LCR
circuit does indeed have strange nonchaotic behaviors.   To distinguish among
the three mechanisms through which SNAs are born, we have examined the manner in
which the maximal Lyapunov exponent and its variance change as a function of
the parameters.  In addition, we have also examined the distribution of local
Lyapunov exponents and found that they take on different characteristics for
different mechanisms.

Given the ubiquity of SNA dynamics in the quasiperiodically driven systems, one of the 
main issues with respect to the observation of SNAs is that this dynamical behaviour occurs 
in a very narrow range of values of the control parameters. While identifying these attractors
from numerical analysis, one may wonder whether they occur due to numerical artifacts and whether 
they may get smeared out if the inherent noise or parameter mismatch is included. For this purpose,
it is important to verify the underlying phenomena experimentally to be sure about the existence 
of the type of transitions to SNAs discussed in this paper. It is here the construction of electronic 
circuits like the one discussed in this manuscript gains physical relevance as an elegant means of experimental verification.

\begin{acknowledgments}
This work has been supported by a Department of Science and Technology,
Government of India sponsored research project. The work of M.~L was carried out under the DAE-BRNS
Raja Ramana Fellowship Programme.
\end{acknowledgments}

\newpage 

\end{document}